\documentclass[useAMs,usenatbib,a4paper]{mn2e}
\usepackage{savesym}
\usepackage{graphicx}
\expandafter\let\csname equation*\endcsname\relax
  \expandafter\let\csname endequation*\endcsname\relax 
\usepackage{subfig}
\usepackage{amsmath}
\usepackage{amssymb}
\usepackage{verbatim}
\usepackage{array}
\usepackage{times}
\usepackage[total={17.8cm,24.0cm},centering]{geometry}

\newcommand{\degree}{\ensuremath{^\circ}}

\newcommand{\be}{\begin{equation}}
\newcommand{\beq}{\begin{equation}}
\newcommand{\ee}{\end{equation}}
\newcommand{\eeq}{\end{equation}}
\newcommand{\eea}{\end{eqnarray}}
\newcommand{\bea}{\begin{eqnarray}}

\newcommand\BV{Brunt-V\"ais\"al\"a\ }

\newcommand\bb[1] { \mbox{\boldmath{$#1$}} }
\newcommand\bcdot{\bb{\cdot}}
\newcommand\del{\bb{\nabla}}

\def\gtsima{$\; \buildrel > \over \sim \;$}
\def\gtsim{\lower.5ex\hbox{\gtsima}}

\def\ltsima{$\; \buildrel < \over \sim \;$}
\def\ltsim{\lower.5ex\hbox{\ltsima}}

\title[Stability and baroclinic rotation in the Sun]{The radiative zone of the Sun and the tachocline: stability of baroclinic patterns of differential rotation}
\author[Andrea Caleo and Steven A. Balbus]{Andrea Caleo \thanks{E-mail:
andrea.caleo@astro.ox.ac.uk} and Steven A. Balbus
\\
Oxford Astrophysics. Denys Wilkinson Building, Keble Road, Oxford, OX1 3RH, United Kingdom}
\begin{document}
\date{}

\pagerange{\pageref{firstpage}--\pageref{lastpage}} \pubyear{2015}

\maketitle

\label{firstpage}

\begin{abstract}
Barotropic rotation and radiative equilibrium are mutually incompatible in stars. The issue is often addressed by allowing for a meridional circulation, but this is not devoid of theoretical complications. Models of rotation in the Sun which maintain strict radiative equilibrium, making use of the observation that the Sun is not in a state of barotropic rotation, have recently been suggested.  To investigate the dynamical behaviour of these solutions, we study the local stability of stratified, weakly magnetized, differentially rotating fluids to non-axisymmetric perturbations.   Finite heat conductivity, kinematic viscosity, and resistivity are present.  The evolution of local embedded perturbations is governed by a set of coupled, ordinary differential equations with time-dependent coefficients.   Two baroclinic models of rotation for the upper radiative zone and tachocline are studied: (i) an interpolation based on helioseismology data, (ii) a theoretical solution directly compatible with radiative equilibrium. The growth of the local Goldreich-Schubert-Fricke instability appears to be suppressed, largely because of the viscosity.   An extensive exploration of wavenumber space is carried out, with and without a magnetic field.  Although we easily find classical local instabilities when they ought formally to be present, for the Sun the analysis reveals neither unstable solutions, nor even solutions featuring a large transient growth.   We have not ruled out larger scale or nonlinear instabilities, nor have we rigorously proven local stability. But rotational configurations in close agreement with observations, generally thought to be vulnerable to the classic local Goldreich-Schubert-Fricke instability, do appear to be locally stable under rather general circumstances. 
\end{abstract}

\begin{keywords}
hydrodynamics  -  instabilities -  stars: rotation -  Sun: helioseismology  -  Sun: interior  -  Sun: rotation 
\end{keywords}

\section{Introduction} \label{sec:intro}
The remarkable development of helioseismology techniques has enabled the study of the internal rotation of the Sun.  The rich and complex findings \citep{Howe2009} were very surprising, as most investigators expected a barotropic pattern of differential rotation fixed on cylinders.  There is no widespread consensus for the explanation of the observations, and the topic is being investigated both by numerical (see e.g. \citealt{BrunMieschToomre2011}) and theoretical analyses (e.g. \citealt{BalbusBonartLatterWeiss2009}). While progress has been made, mainly for our understanding of the convective zone, many issues remain. These include the as yet uncertain physics of the upper radiative zone and the \emph{tachocline}, the relatively thin transition region between the radiative and convective zone.

The well known theorem of Von Zeipel states that a star in uniform or (more generally) barotropic rotation cannot be in radiative equilibrium (see e.g. \citealt{Schwarzschild1956}).  The classic resolution is that residual thermal energy imbalance is compensated by means of a mean velocity flow, the Eddington-Sweet circulation. However, the inclusion of meridional circulation in the equations of stellar structure is ineffective where the entropy gradient vanishes, and there are further complications originating from the azimuthal component of the Euler equation of motion (see \citealt{Caleo2015}, hereafter CBP15, and \citealt{Tassoul2000} for a discussion).  One is then compelled to consider non-steady models or to introduce a well-chosen background magnetic field. Building a simple, comprehensive model for rotating stars, when such meridional circulation is included is not straightforward. Both for these reasons, and because baroclinic rotation does not preclude strict radiative equilibrium, we have calculated an explicit diffusive radiative equilibrium model for the upper radiative zone and tachocline (CBP15). The model is time-steady and features a small baroclinic deviation from uniform rotation.   To be viable, however, such models must be stable.  Are these solutions actually stable?  In this paper, we begin an investigation of this question, by studying the local stability behaviour of such configurations and, more generally, of models compatible with the helioseismology data. 

The literature on the local stability of differential rotation, which is also relevant to the study of accretion discs and accretion flows, is vast.  However, the analysis is difficult to treat in its full generality, i.e. with the inclusion of all the relevant diffusive terms (thermal diffusion, viscosity and resistivity) and allowing for non-axisymmetric perturbations. In this paper, we develop a general diffusive treatment in a local context, deriving a set of equations for the time-evolution of embedded local linear perturbations under conditions in which the equilibrium is both shearing and stratified.    This requires a Lagrangian approach, using coordinates tied to the background flow.  We then apply the equations to the upper radiative zone of the Sun. This extends the study of stability to axisymmetric perturbations by \citet{MenouBalbus2004}, hereafter MBS04, by considering non-axisymmetric modes.   We apply these techniques to two patterns of rotation that are compatible with helioseismology data. 

An outline of the paper is as follows. Section 1.1 summarizes the current knowledge on the local stability of differentially rotating fluids and motivates the need of a general treatment which includes the presence of finite diffusivities and non-axisymmetric perturbations. Section 2 is a derivation of the equations for the evolution of a perturbation in this general case. In section 3, we introduce two models of differential rotation for the upper radiative zone. In section 4 we revisit the Goldreich-Schubert-Fricke (hereafter GSF) instability (\citealt{GoldreichSchubert}, hereafter GS67) and discuss its onset for the two models of rotation. In section 5 we present a study of a large variety of perturbation solutions, with initial wavenumbers selected both systematically and (to minimise bias) stochastically across the space of parameters. Section 6 summarizes our results.

\subsection{Previous results on the stability of differentially rotating stars}  \label{sec:intr1}

We use both standard spherical coordinates $(r, \phi, \theta)$ as well as cylindrical coordinates $(R, \phi, z)$.
The background rotation is azimuthally symmetric, but otherwise arbitrary:  $\Omega = \Omega (r, \theta)$ or $\Omega = \Omega(R, z)$.

\subsubsection{Axisymmetric perturbations} \label{sec:intr11}
The stability conditions for adiabatic, axisymmetric perturbations in a non-magnetic star are
a classic result from the linear theory of differentially rotating stars, historically attributed to Solberg and H\o iland (e.g. \citealt{Tassoul2000}).   The Solberg-Hoiland criteria state that such systems are stable if:
\begin{equation}  \label{Hoiland1}
N^2 + \frac{1}{R^3} \frac{\partial l^2}{\partial  R} > 0 ,
\end{equation}
\begin{equation}  \label{Hoiland2}
(- \frac{\partial P}{\partial z}) (\frac{\partial l^2}{\partial R} \frac{\partial \sigma}{\partial z} - \frac{\partial l^2}{\partial z} \frac{\partial \sigma}{\partial R}) > 0 ,
\end{equation}
where $N^2$ is the usual \BV frequency 
\begin{equation}  \label{Brunt–Vaisala}
N^2 = - \frac{1}{\gamma \rho} \frac{d P}{d r} \frac{d \sigma}{dr} ,
\end{equation}
and $\sigma = \ln{(P \rho^{-\gamma})}$ is a dimensionless entropy variable.  Finally,
$\gamma$ is the adiabatic index, $\rho$ the mass density in the star, $P$ the pressure, and $l$ the specific angular momentum:
\begin{equation}  \label{Brunt–Vaisala}
l = \Omega R^2 .
\end{equation}

An important modification of this result comes from introducing a finite thermal diffusivity. The destabilising effect of thermal diffusion was studied by GS67 and \citet{Fricke1968}, who considered axisymmetric perturbations in a non-magnetic star with a finite thermal diffusion coefficient $\xi_{\text{rad}}$ and kinematic viscosity $\nu$. The authors found that if the heat leakage due to thermal diffusion from a perturbed fluid element is sufficiently rapid, the restoring effect of buoyancy is lost.   It was found that when
\begin{equation} \label{GSlimit}
\nu / \xi_{\text{rad}} \rightarrow 0,
\end{equation}
there are always unstable modes unless
\begin{equation} \label{GSCondition1}
\frac{\partial (\Omega R^2)}{\partial R} > 0
\end{equation}
and
\begin{equation} \label{GSCondition2}
\frac{\partial \Omega}{\partial z} = 0,
\end{equation}
i.e., the specific angular momentum must increase outwards, as per the well-known Rayleigh criterion, and the angular velocity must be fixed on cylinders. 

How efficient would the GSF instability be in redistributing the angular momentum in a star that doesn't comply with equation \eqref{GSCondition2}? The answer to this question is not simple, even in the linear regime. The growth time-scale of the instability is relatively long, generally of order the Kelvin-Helmoltz time-scale \citep{Kippenhahn1969}, and possibly longer 
(\citealt{JamesKahn1971}, \citealt{Kippenhahnetal1980}). A small compositional gradient could render the fluid stable \citep{KnoblochSpruit1983}.   Finally, a subtle feature of the analysis by GS67, though noted early on (e.g. \citealt{Acheson1978}, \citealt{KnoblochSpruit1982}, and MBS04), is often ignored.  Equations \eqref{GSCondition1} and \eqref{GSCondition2} are strictly correct only when $\nu / \xi_{\text{rad}}$ is sufficiently small.   Small compared to what?  A careful analysis shows that being small compared to unity is not sufficient: $\nu / \xi_{\text{rad}}$ must be small compared with the squared ratio of  $\Omega$ over the \BV frequency.  This condition may not be strictly valid in much of the radiative zone.   It is therefore necessary to retain the viscosity $\nu$ when studying the behaviour of these local perturbations.  We discuss this issue in section \ref{sec:GS}.   While other hydrodynamic instabilities have been investigated in the years following (e.g. \citealt{Heger2000}), the GSF instability deserves particular attention in stellar radiative zones.  In fact, \citet{KnoblochSpruit1982} argued that the GSF instability is one of the few hydrodynamic processes that can transport angular momentum across surfaces of constant pressure.

As a remnant of their formation, the Sun and the other stars are expected to have an internal magnetic field of indeterminate strength.   The diffusion time-scale of such a fossil magnetic field is very long (see e.g. \citealt{Parker1979} and \citealt{Mestel1999}). Developments in the last decades, stemming from the studies of accretion disc stability, have made it clear that consideration of even weak magnetic fields is often essential for an understanding of the dynamics of shearing fluids (see e.g. the magneto-rotational instability (MRI), \citealt{BalbusHawley1991}). The effects of toroidal magnetic fields in the context of differentially rotating stars had been discussed by \citet{Acheson1978}, who noted that the presence of such weak fields can change the results by GS67.  \citet{Balbus1995} derived the dispersion relation for linear axisymmetric perturbations in a weakly magnetized non-diffusive star. For even very weak background magnetic fields, the conditions for stability are independent of the magnetic field:   
\begin{equation}\label{BalbusI}
N^2 +{\partial\Omega^2\over \partial \ln R} > 0,
\end{equation}
\begin{equation}  \label{Balbus2}
(- \frac{\partial P}{\partial z}) (\frac{\partial \Omega^2}{\partial R} \frac{\partial \sigma}{\partial z} - \frac{\partial \Omega^2}{\partial z} \frac{\partial \sigma}{\partial R}) > 0 ,
\end{equation}
i.e., the same as equations \eqref{Hoiland1} and  \eqref{Hoiland2} but with the angular velocity replacing the specific angular momentum.  The range of magnetic field strength for the validity of the stability criteria \eqref{BalbusI} and \eqref{Balbus2} depends on the diffusion coefficients (see section 2.4 of \citealt{Balbus1995} for a discussion). For typical solar values, the criterion is valid for $B > 10^2$ G. 

These considerations prompted MBS04 to conduct a more general axisymmetric analysis in magnetised stars, including thermal diffusivity $\xi_{\text{rad}}$, viscosity $\nu$, and magnetic resistivity $\eta$. They derived a fifth-order dispersion relation and discussed several necessary conditions for stability in the case in which one of the three diffusivities is zero, recovering the conditions we have described here. They then applied the full dispersion relation to the upper radiative zone of the Sun, finding that the addition of a third, weak diffusivity is often able to stabilise an unstable double-diffusive situation. The radiative zone of the Sun may be subject to unstable modes if moderate or strong radial gradients of angular velocity were present. This is not the case, at least in the bulk of the radiative zone, which is found to be in near-uniform rotation (see section \ref{sec:stabilityofthesun}). \citet{MenouLeMer2006} have then surveyed the axisymmetric stability of the early (faster rotating) Sun and showed that it may have been more prone to rotational instabilities than the current Sun. They found that MHD modes are most likely more efficient at transporting angular momentum than the hydrodynamic instabilities in stellar interiors.

\subsubsection{Non-axisymmetric perturbations} \label{sec:non-axisymmetric}
Studies of the stability of differentially rotating stars is often restricted to axisymmetric perturbations. This is primarily due to the complications associated with the behaviour of non-axisymmetric fluid displacements, which in general cannot be described by a local plane wave dispersion relation.

A study of the local linear behaviour of three-dimensional fluid displacements in a shearing and stratified background medium was conducted by \citet{BalbusSchaan2012} (the classic ``shearing-sheet'' reference is \citealt{GoldreichLyndenBell1964}). They make use of Lagrangian cylindrical coordinates $(R', \phi', z', t')$, locally comoving with the fluid and related to the Eulerian coordinates by:
\begin{equation}  
R' = R, \qquad \phi' = \phi - \Omega t, \qquad z' = z, \qquad t' = t ,
\end{equation}
with $\Omega=\Omega (R, z)$.  In the WKB limit, when written in Lagrangian coordinates the perturbations have the familiar plane wave spatial dependence:
\begin{equation}  \label{WKBLagrange}
\exp[i(k_R' R' + m \phi' + k_z' z')].
\end{equation}
In these (but not in standard Eulerian) coordinates, the components of the wave vector $\bb k'$ are constants. 

The partial derivatives with respect to the Eulerian coordinates $(R, \phi, z, t)$ are related to the Lagrangian coordinates by:
\begin{equation}  
\frac{\partial}{\partial R} = \frac{\partial}{\partial R'} - t \frac{\partial \Omega}{\partial R} \frac{\partial}{\partial \phi'} ,
\end{equation}
\begin{equation}  
\frac{\partial}{\partial \phi} = \frac{\partial}{\partial \phi'},
\end{equation}
\begin{equation}  
\frac{\partial}{\partial z} = \frac{\partial}{\partial z'} - t \frac{\partial \Omega}{\partial z} \frac{\partial}{\partial \phi'} ,
\end{equation}
\begin{equation}  
\frac{\partial}{\partial t} = \frac{\partial}{\partial t'} - \Omega \frac{\partial }{\partial \phi'} .
\end{equation}
This is equivalent to introducing time dependent Eulerian waveumbers,
\begin{equation}  \label{EulerianWavenumbers}
k_R (t) = k_R' - m t \frac{\partial \Omega}{\partial R}, \qquad k_z (t) = k_z' - m t \frac{\partial \Omega}{\partial z},
\end{equation}
with $m$ unchanged.   (In this way, it is possible to formulate the problem entirely in Eulerian
coordinates, though in our view it is less natural.)  
Since the Eulerian wavenumbers depend on time, the coefficients in the governing evolutionary fluid equations must depend on time as well, and simple harmonic wave solutions of the form 
$e^{i \omega t}$ do not exist.

The wave vector $\bb k$ is constant either when $m = 0$ (axisymmetry), or when 
\begin{equation}  
\frac{\partial \Omega}{\partial R} = 0 \qquad \text{and \ \ \ \ \  \ \ \ } \frac{\partial \Omega}{\partial z} = 0 ,
\end{equation}
i.e., the case of uniform rotation. In these two instances, a dispersion relation for the evolution of the perturbations can be derived. \citet{BalbusSchaan2012} discuss the dispersion relation for non-axisymmetric displacements in the case of uniform rotation.  They note that in some instances, ``\emph{the pure $m$ modes are more unstable than are modes contaminated by poloidal wavenumber components}''.  For convectively unstable flows it is not unusual to find that the axisymmetric perturbations are stable, whereas high $m$ modes are not. It is therefore of great importance to consider the full 3-dimensional case.

Several studies (e.g. \citealt{Acheson1978}, \citealt{Masada2007},and \citealt{KaganWheeler2014})
neglect the time-dependence of $k_R$ and $k_z$ in the nonaxisymmetric case.   This restricts their range of validity.  The latter two papers explicitly note that the analysis is valid only when the axisymmetric component of the wave vector is small enough not to be significantly sheared with time.   However, it is unclear how this approximation is included in the analysis, and the actual {\em dynamics} (as opposed to the
rotational kinematics) may be correct only to zeroth order in $m$, i.e. the axisymmetric limit.  In particular, \citet{KaganWheeler2014} (see also \citealt{ParfreyMenou2007}) study the growth rate of the MRI in the upper radiative and convective regions of the Sun and suggest that the unstable MRI modes may play a dominant role in the generation of the toroidal magnetic field near the tachocline.  However, this conclusion is based upon the behaviour of their fastest growing modes, which are, in fact, fully non-axisymmetric.    This rich problem merits a fully self-consistent nonaxisymmetric local analysis.  

\section{Local linear growth of perturbations in a shearing fluid}
We present here the equations for the local linear growth of incompressible WKB perturbations in a weakly magnetized, shearing fluid in the triple-diffusive case for a generic Lagrangian wave vector $\bb k_R' = (k_R', k_\phi', k_z')$.  The axisymmetric limit of this is presented in MBS04, while the adiabatic limit is presented in \citet{BalbusSchaan2012}.  The azimuthal component of the wave vector, expressed henceforth as $k_\phi'$, is related to the notation of section \ref{sec:non-axisymmetric} by $k_\phi' = m / R$. 

\subsection{The fluid equations}
The governing equations are:
\begin{equation}  \label{EqCont}
\frac{\partial \rho}{\partial t} + \bb \nabla \bcdot (\rho \bb v) = 0, 
\end{equation}
\begin{equation}  \label{EqMotion}
\rho \frac{D \bb v}{D t} + \bb \nabla \Big( P + \frac{B^2}{8 \pi} \Big) - \frac{1}{4 \pi} (\bb B \bcdot \bb \nabla) \bb B - \rho \bb g - \rho \nu \nabla^2 \bb v = 0, 
\end{equation}
\begin{equation}  \label{EqMagneticField}
\frac{\partial \bb B}{\partial t} - \bb \nabla \times ( \bb v \times \bb B) - \eta \nabla^2 \bb B = 0 ,
\end{equation}
\begin{equation}  \label{Entropy}
\frac{P}{\gamma - 1} \frac{D \sigma}{D t} - \chi \nabla^2 T = 0 , 
\end{equation}
where $\bb v$ is the velocity of the fluid, $\rho$ its density, $P$ its pressure, $\bb B$ the magnetic field, ${\bf g}$ the gravitational field, $\nu$ the kinematic viscosity, $\eta$ the resistivity, $\sigma = \log{(P \rho^{-\gamma})}$, $\gamma$ the adiabatic index, $\chi$ the heat conductivity, and $T$ the temperature. The Lagrangian derivative $D/Dt$ in its Eulerian form is as usual: 
\begin{equation}  \label{LagrangianDerivative}
\frac{D}{Dt} = \frac{\partial}{\partial t} + (\bb v \bcdot \bb \nabla) .
\end{equation}
As in MBS04, we neglect the spatial dependence of the diffusion coefficients, which is appropriate for a local WKB analysis. We also neglect the resistive and viscous dissipation terms in equation \eqref{Entropy}, as they appear only at higher order; see Appendix A1 of MBS04.

We consider Eulerian perturbations (denoted by a prefix $\delta$) of a background medium. The perturbations have a local WKB plane wave spatial dependence in the Lagrangian frame, as described in section \ref{sec:non-axisymmetric}, with Lagrangian wave vector $\bb k' = (k_R', k_\phi', k_z')$ and corresponding Eulerian wave vector $\bb k(t) = (k_R(t), k_\phi, k_z(t))$ where $k_R(t)$ and $k_z(t)$ are given by equation \eqref{EulerianWavenumbers}. The equilibrium state rotation is given by $\bb \Omega = \Omega (R, z) \hat z$ along the $z$-axis. We neglect any bulk circulation in the star. We also neglect the effects our perturbations have on the gravitational potential of the star (Cowling approximation).  We make the weak field approximation and assume that the magnetic field plays no role in the background equilibrium state, but it can still be important for the evolution of perturbations with large wavenumbers.  We will explicitly state whenever we neglect a magnetic term when equations \eqref{EqMotion} - \eqref{Entropy} are perturbed.

We restrict nearly incompressible modes adopting a modified Boussinesq approximation (\citealt{Spiegel1960}, \citealt{KunduCohen}).   This involves replacing the continuity equation \eqref{EqCont} with the simpler condition:
\begin{equation}  \label{Boussinesq}
\bb \nabla \bcdot \delta \bb v = 0,
\end{equation}
and neglecting perturbations in the density in equations \eqref{EqMotion} - \eqref{Entropy}, except when $\rho$ is multiplied by the gravity $\bb g$, and (the ``modification'') in the entropy equation where it
dominates over pressure perturbations $\delta P$.  In our problem, equation \eqref{Boussinesq} translates to
\begin{equation}  \label{Boussinesq2}
\bb k \bcdot \delta \bb v = 0.
\end{equation}
In this equation and in those that follow, it is understood that $\bb k$ always refers to the Eulerian, time-dependent wave vector.
   
We also of course require the magnetic field to be divergence-free:
\begin{equation}  \label{divB}
\bb k \bcdot \delta \bb B = 0,
\end{equation}
a condition already ensured by (\ref{EqMagneticField}) if it is imposed at $t=0$.   

\subsection{First-order expansion of \eqref{EqMotion}} \label{secEqMotion}
With a linear perturbation added to the velocity field, we set $\bb v = \Omega R \hat \phi + \delta \bb v$ in equation \eqref{EqMotion}.  The leading order linear terms are given by
\begin{equation}  
	\begin{aligned}
	&\delta\left( \frac{D \bb v}{D t}\right) = & & \Big( \frac{D \delta v_R}{D t} - 2 \Omega \delta v_\phi \Big) \hat R \ + & \\
	&	& +	& \Big( \frac{D \delta v_\phi}{D t} + \Omega \delta v_R + \delta \bb v \bcdot \bb \nabla (\Omega R) \Big) \hat \phi \ + & \\
	&	&	+	& \Big( \frac{D \delta v_z}{D t} \Big) \hat z .
	\end{aligned}
\end{equation}
To express the first-order approximation of the second term of equation \eqref{EqMotion}, we define the total (gas + magnetic) pressure in the plasma as:
\begin{equation}
P_t = P + \frac{B^2}{8 \pi} .
\end{equation}
The result is:
\begin{equation}
 \delta\left(\bb \nabla P_t\right) \cong \Big(i \frac{k_R}{\rho} \delta P_t \Big) \hat R + \Big( i \frac{k_\phi}{\rho} \delta P_t \Big) \hat \phi +	\Big(i \frac{k_z}{\rho} \delta P_t \Big) \hat z .
\end{equation}
Assuming the equilibrium magnetic field and its spatial derivatives to be small, we neglect the $(\delta \bb B \bcdot \bb \nabla) \bb B$ term in the first-order expression of the third term of equation \eqref{EqMotion} relative to spatial gradients of $\delta\bb{B}$. The result is:
\begin{equation}
\frac{1}{4 \pi} ( \bb B \bcdot \bb \nabla ) \delta \bb B \cong \frac{1}{4 \pi} i (\bb k \bcdot \bb B) (\delta B_R \hat R + \delta B_\phi \hat \phi + \delta B_z \hat z)
\end{equation}
We note that the azimuthal component of the background magnetic field $B_\phi$ is not independent of time, but satisfies the induction equation:
\begin{equation}
\frac{\partial B_\phi}{\partial t} = R (\bb B \bcdot \bb \nabla) \Omega,
\end{equation}
which gives
\begin{equation} \label{Bphiwithtime}
B_\phi = B_\phi' + t R (\bb B \bcdot \bb \nabla) \Omega ,
\end{equation}
where $B_\phi'$ is the initial azimuthal component of the magnetic field. Computing the dot product $\bb k \bcdot \bb B$ by means of equations \eqref{EulerianWavenumbers} and \eqref{Bphiwithtime}, we obtain $\bb k (t) \bcdot \bb B(t) = \bb k' \bcdot \bb B'$. The quantity $\bb k \bcdot \bb B$ is thus independent of time.

We adopt the standard Cowling approximation and neglect the changes to the gravitational potential caused by linear perturbations.   We retain the buoyancy term in $\delta \rho\,  \bb g$ when perturbing equation \eqref{EqMotion}.    Assuming hydrostatic equilibrium and axisymmetry for the background state, we have:
\begin{equation} 
\bb g = \frac{1}{\rho} \frac{\partial P_t}{\partial R} \hat R + \frac{1}{\rho} \frac{\partial P_t}{\partial z} \hat z.
\end{equation}
Therefore:
\begin{equation} 
\delta (\rho \bb g) = (\delta \rho) \bb g = \Big( \frac{\delta \rho}{\rho} \frac{\partial P_t}{\partial R} \hat R + \frac{\delta \rho}{\rho} \frac{\partial P_t}{\partial z} \hat z \Big) . 
\end{equation}
Finally, the first-order approximation of the last term of equation \eqref{EqMotion} is:
\begin{equation}
- \rho \nu \nabla^2 \bb v \simeq  \rho \nu k^2 (\delta v_R \hat R + \delta v_\phi \hat \phi + \delta v_z \hat z) .
\end{equation}
The resulting equations for the three components of equation \eqref{EqMotion} are:

\begin{equation} \label{PertMot1}
 \frac{D \delta v_R}{D t} = 2 \Omega \delta v_\phi + \frac{\delta \rho}{\rho^2} \frac{\partial P_t}{\partial R} -  \frac{i k_R}{\rho} \delta P_t + \frac{i (\bb k \bcdot \bb B)}{4 \pi \rho} \delta B_R - k^2 \nu \delta v_R  ,
\end{equation}
\begin{equation}\label{PertMot2}
\frac{D \delta v_\phi}{D t} = - \Omega \delta v_R - \delta \bb v \bcdot \bb \nabla(\Omega R) - \frac{i k_\phi}{\rho} \delta P_t + \frac{i (\bb k \bcdot \bb B)}{4 \pi \rho} \delta B_\phi - k^2 \nu \delta v_\phi ,
\end{equation}
\begin{equation} \label{PertMot3}
\frac{D \delta v_z}{D t} = \frac{\delta \rho}{\rho^2} \frac{\partial P_t}{\partial z} -  \frac{i k_z}{\rho} \delta P_t  +\frac{i (\bb k \bcdot \bb B)}{4 \pi \rho} \delta B_z - k^2 \nu \delta v_z .
\end{equation}

\subsection{First-order expansion of \eqref{EqMagneticField}}
In the first-order expansion of equation \eqref{EqMagneticField}, we neglect the term $(\delta \bb v \bcdot \bb \nabla)\bb B$ by the weak field and WKB assumptions. Proceeding as above, the components of the perturbed equation are found to be:
\begin{equation} \label{PertMag1}
\frac{D \delta B_R}{D t} - i (\bb k \bcdot \bb B) \delta v_R + \eta k^2 \delta B_R = 0 ,
\end{equation}
\begin{equation} \label{PertMag2}
\frac{D \delta B_\phi}{D t} - i (\bb k \bcdot \bb B) \delta v_\phi - R \big( (\delta \bb B \bcdot \bb \nabla) \Omega \big) + \eta k^2 \delta B_\phi = 0 ,
\end{equation}
\begin{equation} \label{PertMag3}
\frac{D \delta B_z}{D t} - i (\bb k \bcdot \bb B) \delta v_z + \eta k^2 \delta B_z = 0  .
\end{equation}

\subsection{First-order expansion of \eqref{Entropy}}
To compute the first-order expansion of equation \eqref{Entropy}, we make the standard assumption that the displaced fluid element is in near pressure equilibrium with its surrounding, because the time-scale to reach such equilibrium is much shorter than any other relevant time-scale. 
\begin{equation}
\delta \sigma = \delta \log{(P \rho^{-\gamma})} = \frac{\delta P}{P} - \gamma \frac{\delta \rho}{\rho} \simeq - \gamma \frac{\delta \rho}{\rho} .
\end{equation}
We also assume a perfect gas equation of state,
\begin{equation}
P = \frac{\rho}{\mu m_P} k_B T ,
\end{equation}
where $\mu$ is the average chemical weight, $m_P$ is the mass of the proton, and $k_B$ is Boltzmann's constant. Assuming a uniform chemical composition, we relate the perturbations in the pressure, density, and temperature by:
\begin{equation}
\frac{\delta T}{T} = \frac{\delta P}{P} - \frac{\delta \rho}{\rho} \simeq - \frac{\delta \rho}{\rho} . 
\end{equation}
(We note the retention of the $\delta\rho$ term, in accordance with our modified Boussinesq limit.)
Proceeding as in section (\ref{secEqMotion}), the perturbed entropy equation is found to be:
\begin{equation} \label{PertEntropy}
\frac{D}{Dt}\Big( \frac{\delta \rho}{\rho} \Big) - \frac{1}{\gamma} \Big( \delta v_R \frac{\partial \sigma}{\partial R} + \delta v_z \frac{\partial \sigma}{\partial z} \Big) + \frac{\gamma - 1}{\gamma} \chi k^2 \frac{T}{P} \frac{\delta \rho}{\rho} = 0 .
\end{equation}

\subsection{Reduction to real, coupled equations}
We wish to reduce the perturbed equations to a set of real, coupled ordinary differential equations.  By multiplying equations \eqref{PertMag1}, \eqref{PertMag2}, and \eqref{PertMag3} by $i$, the magnetic field perturbation $\delta \bb B$ always appears with a factor if $i$ in the perturbed equations.  This is equivalent to a simple change of phase by $\pi / 2$. To obtain a set of real equations, we work with the variable
\begin{equation}
\delta \bb C = i \delta \bb B ,
\end{equation}
in preference to $\delta \bb B$ itself.

We make use of equations \eqref{Boussinesq} and \eqref{divB} to replace $\delta v_\phi$ and $\delta C_\phi$:
\begin{equation} \label{deltavphi}
\delta v_\phi = - \frac{1}{k_\phi}(k_R \delta v_R + k_z \delta v_z) ,
\end{equation}
\begin{equation}\label{deltaceephi}
\delta C_\phi = - \frac{1}{k_\phi}(k_R \delta C_R + k_z \delta C_z).
\end{equation}
It is also necessary to compute the Lagrangian time derivative of equation \eqref{deltavphi}, since the term $D \delta v_\phi / Dt$ appears in equation \eqref{PertMot2}. The result is:
\begin{equation} 
\frac{D \delta v_\phi}{D t} = - \frac{1}{k_\phi} \Big( k_R \frac{D \delta v_R}{D t} + k_z \frac{D \delta v_z}{D t} \Big) + R (\delta \bb v \bcdot \bb \nabla) \Omega .
\end{equation}

\subsection{Governing equations}

Our strategy is to eliminate $\delta v_\phi$ and $\delta C_\phi$ in equations \eqref{PertMot1}--\eqref{PertMag3} and\eqref{PertEntropy}).
We may then solve for
\begin{equation} 
i \frac{1}{\rho} \delta P_t
\end{equation}
in equation \eqref{PertMot2} and substitute this into equations \eqref{PertMot1} and \eqref{PertMot3}. 
Finally, we rearrange equations \eqref{PertMot1} and \eqref{PertMot3} to isolate the Lagrangian derivatives of $\delta v_R$ and $\delta v_z$. The procedure is quite lengthy, but entirely straightforward. 
Defining the Alfv\'en velocity:
\begin{equation} 
\bb v_A = \frac{\bb B}{\sqrt{4 \pi \rho}} ,
\end{equation}
 the final form of the equations are:
\begin{equation} \label{Eq1}
\begin{aligned}
\frac{D \delta v_R}{Dt} & - \frac{\bb k \bcdot \bb v_A}{\sqrt{4 \pi \rho}} \delta C_R + 2 \Omega \frac{k_\phi^2 + k_z^2}{k_\phi k^2} (k_R \delta v_R + k_z \delta v_z) - \\
& - 2 \frac{k_\phi k_R}{k^2} (\delta \bb v \bcdot \bb \nabla) (\Omega R) + \frac{\delta \rho}{\rho^2} \Big( \frac{k_z^2}{k^2} \widetilde D P - \frac{k_\phi^2}{k^2} \frac{\partial P}{\partial R}  \Big) + \\ 
& + \nu k^2 \delta v_R = 0 ,
\end{aligned}
\end{equation}
\begin{equation} \label{Eq2}
\begin{aligned}
\frac{D \delta v_z}{Dt} & - \frac{\bb k \bcdot \bb v_A}{\sqrt{4 \pi \rho}} \delta C_z - 2 \Omega \frac{k_R k_z}{k_\phi k^2} (k_R \delta v_R + k_z \delta v_z) - \\
& - 2 \frac{k_\phi k_z}{k^2} (\delta \bb v \bcdot \bb \nabla) (\Omega R) - \frac{\delta \rho}{\rho^2} \Big( \frac{k_R k_z}{k^2} \widetilde D P + \frac{k_\phi^2}{k^2} \frac{\partial P}{\partial z}  \Big) + \\ 
& + \nu k^2 \delta v_z = 0 ,
\end{aligned}
\end{equation}
\begin{equation} \label{Eq3}
\frac{D \delta C_R}{Dt} + \eta k^2 \delta C_R + \sqrt{4 \pi \rho} (\bb k \bcdot \bb v_A) \delta v_R = 0 ,
\end{equation}
\begin{equation} \label{Eq4}
\frac{D \delta C_z}{Dt} + \eta k^2 \delta C_z + \sqrt{4 \pi \rho} (\bb k \bcdot \bb v_A) \delta v_z = 0 ,
\end{equation}
\begin{equation} \label{Eq5}
\frac{D}{Dt} \Big( \frac{\delta \rho}{\rho} \Big) - \frac{1}{\gamma} \Big( \delta v_R \frac{\partial \sigma}{\partial R} + \delta v_z \frac{\partial \sigma}{\partial z} \Big) + \frac{\gamma - 1}{\gamma} \chi k^2 \frac{T}{P} \frac{\delta \rho}{\rho} = 0.
\end{equation}
In these equations, we have  used the $\widetilde D$ operator \citep{Balbus1995}:
\begin{equation} 
\widetilde D = \frac{k_R}{k_z} \frac{\partial}{\partial z} - \frac{\partial}{\partial R},
\end{equation}
and the coordinates of the Eulerian wave vector $\bb k$ are given by:
\begin{equation}  \label{kRoft}
k_R (t) = k_{R0} - k_{\phi 0} R t \frac{\partial \Omega}{\partial R}, 
\end{equation}
\begin{equation} 
{m\over R}\equiv k_\phi  = k_{\phi0} = {\rm constant}, 
\end{equation}
\begin{equation}    \label{kzoft}
k_z (t) = k_{z0} - k_{\phi0} R t \frac{\partial \Omega}{\partial z} ,
\end{equation}
where $k_{R0}$, $k_{\phi0}$, and $k_{z0}$ are the initial values of the wave vector components (corresponding to $k_R'$, $k_\phi'$, and $k_z'$ in the notation of the previous sections).
Finally, the azimuthal components of $\delta \bb v$ and $\delta \bb B$ are given by equations (\ref{deltavphi})
and (\ref{deltaceephi}).
Equations \eqref{Eq1} - \eqref{Eq5} constitute a set of five ordinary differential equations with time-dependent coefficients in the variables $\delta v_R$, $\delta v_z$, $\delta C_R$, $\delta C_z$, and $\delta \rho$. 

\section{The rotation of the upper radiative zone of the Sun}  \label{sec:stabilityofthesun}
\subsection {Model parameters}

We use equations \eqref{Eq1} - \eqref{Eq5} to analyse the local stability of the upper radiative zone of the Sun.  The background state is given by a standard, non-rotating solar model (\citealt{BahcallSerenelliBasu2005}). We then extract local values for the background $P$, $\rho$, and $T$ and calculate the gravitational field in the star from the mass distribution. The background state in effect satisfies our equation \eqref{EqMotion} with $\bb v = 0$ and $\bb B = 0$.   Just as with the magnetic field, the rotation does not
influence hydrostatic equilibrium at a significant level, but it is important for the behaviour of perturbations.  

\begin{figure}
	\centering
	\includegraphics[width=0.5\textwidth, clip=true, trim=0cm 0cm 0cm 0cm]{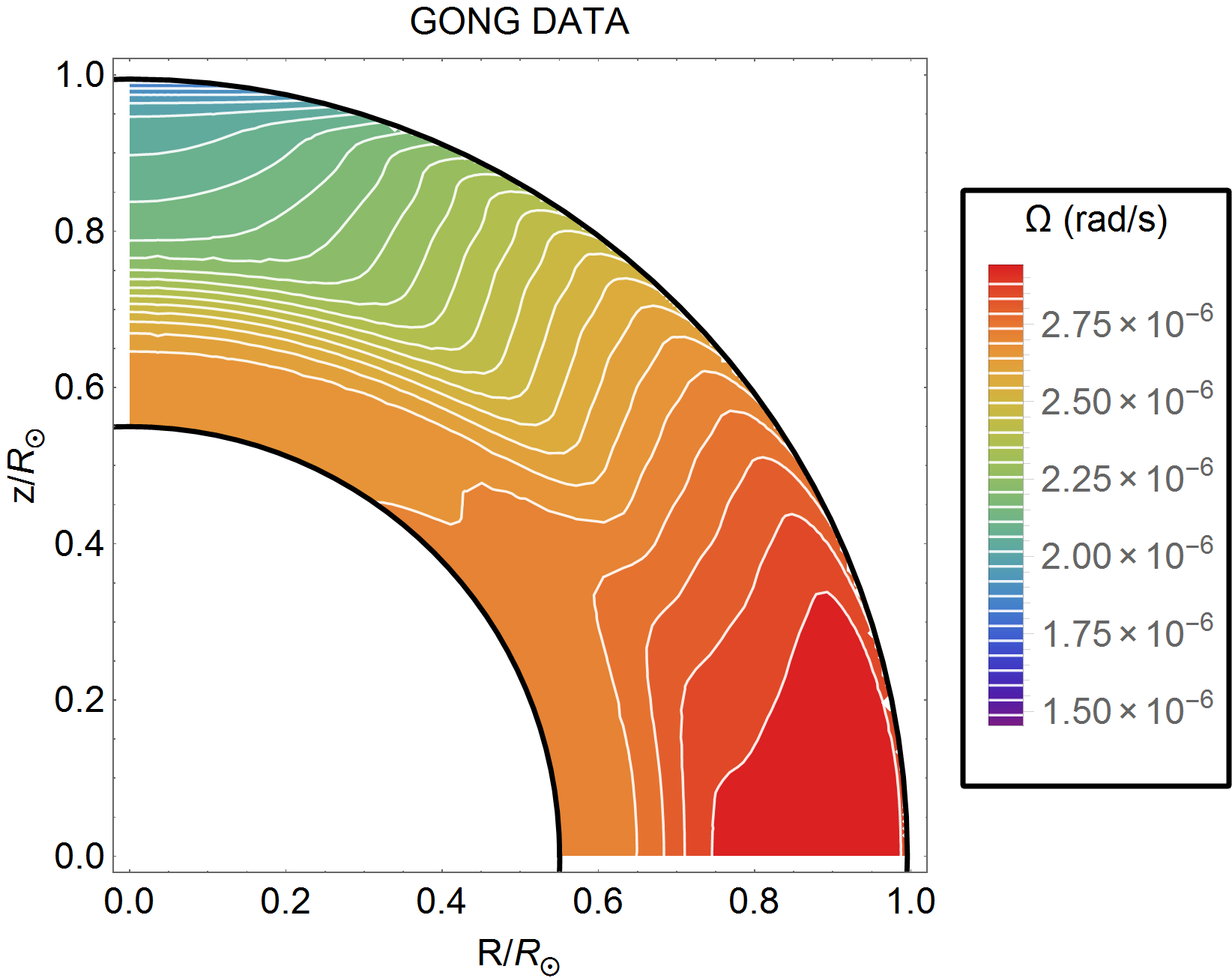}					
	\caption{\label{figGONG}Isocontours of the angular velocity $\Omega(r, \theta)$ for $0.55 R_\odot < r < R_\odot$ (GONG data, \citealt{GONGData}).}
\end{figure}

Observational constraints on the angular velocity are provided by helioseismology, which shows a pattern of approximately uniform rotation throughout much of the radiative zone, but with significant shear near the convective boundary.  The uncertainty on $\Omega$ depends on latitude, with values at the poles being relatively more inaccurate.  It is estimated to be of order $10\%$ at the depth of the radiative zone (R. Howe, private communication; see also \citealt{EffDarwich2013}). The situation is more difficult to assess in the tachocline, which is not particularly well-resolved.  Models exist for the physics of the upper radiative zone yielding rotation curves compatible with the helioseismology data, but neither the models nor observations adhere to the strict GSF local stability criterion of $\Omega$ being fixed on cylinders, $\partial_z \Omega = 0$ (CBP15).  
In sections \ref{sec:GS} and \ref{sec:trials} we investigate the local stability behaviour of two such ``GSF-violating'' patterns of differential rotation with $\partial_z \Omega \ne 0$, which we refer to as models A and B:
\begin{itemize}
\item{Model A is the interpolation of a recent set of helioseismology data by the GONG group \citep{GONGData}. The data and isorotation contours are shown in figure \ref{figGONG}. We approximate $\Omega(r, \theta)$ with a function of the form:
\begin{equation} \label{OmegaFit}
\Omega^2(r, \theta) \cong \Omega_0^2(r) + \Omega_2^2(r) \cos^2{\theta},
\end{equation}
where fifth-order polynomials are used for $\Omega_0^2(r)$ and $\Omega_2^2(r)$. This approximation 
reproduces the helioseismology data very accurately (CBP15). 
While the Sun may or may not be in static radiative equilibrium depending on fine rotational details, this model would, strictly speaking, require some form of circulation or evolution to maintain radiative equilibrium.  This would not affect the analysis of the present paper, which is concerned with timescales much less than Kelvin-Helmholtz. }
\item{Model B is the radiative rotation curve described by CBP15. This curve, which also provides a good fit to the GONG data, is derived by imposing the requirement of {\em exact} radiative equilibrium $\bb \del \bcdot \bb F_{\text{rad}} = 0$ in the radiative zone and in the tachocline. Model B represents an alternative to the standard uniform rotation model with thermal equilibrium maintained by meridional circulation.  It describes a time-steady, circulation-free state of strict radiative equilibrium, both in the bulk of the radiative zone (where the differential rotation is tiny) and in the tachocline (where it is significant).   Model A and model B are both compatible with the data at the current level of accuracy.}
\end{itemize}

Models A and B are numerically similar, although their physical interpretation is different; see CBP2015 for a quantitative comparison. We report here just the angular velocity and its first derivatives at $r = 0.7 R_\odot$ and $\theta = 45 \degree$ for the two models:
\begin{equation} 
\Omega_A = 2.7 \times 10^{-6} \text{rad s}^{-1}, \qquad \Omega_B = 2.7 \times 10^{-6} \text{rad s}^{-1},
\end{equation}
\begin{equation} 
\Big( \frac{\partial \ln \Omega}{\partial \ln R} \Big)_A = - 0.11 , \qquad \Big( \frac{\partial \ln \Omega}{\partial \ln R} \Big)_B = - 0.15,
\end{equation}
\begin{equation} 
\Big( \frac{\partial \ln \Omega}{\partial \ln z} \Big)_A = - 0.24 , \qquad \Big( \frac{\partial \ln \Omega}{\partial \ln z} \Big)_B = - 0.26.
\end{equation}
For comparison, $(N/\Omega)^2 \cong 1.6 \times 10^5$ in both models\footnote{This shows that the stability criterion \eqref{BalbusI} is not violated at this depth and latitude. We note that since there are locations near the tachocline with $\partial \ln \Omega^2 / \partial \ln R < 0$, there must be a region in which $N^2$ goes through zero and equation \eqref{BalbusI} is not satisfied. However, this region is extremely narrow in the current standard Solar model and becomes immediately convectively unstable.}.

The diffusivities are estimated in the same manner as MBS04.  We denote by $\chi$ the coefficient so denoted by the same letter by MBS04. This differs from $\chi$ as used in GS67, even dimensionally. The latter is related to the coefficient $\xi_{\text{rad}}$ by MBS04 by: 
\begin{equation}
\xi_{\text{rad}} = (\gamma - 1) \chi_{GS},
\end{equation}
where $\chi_{GS}$ is the quantity denoted as $\chi$ by GS67. Dimensionally, $[\chi] = M L T^{-3} K^{-1}$, $[\xi_{\text{rad}}] = [\chi_{GS}] = L^2 T^{-1}$, where $M, L, T, K$ respectively represent mass, length, time, and temperature.

The thermal diffusion coefficients in the radiative zone of the Sun are given by (see e.g. \citealt{Schwarzschild1956}):
\begin{equation}
\chi = \frac{16 T^3 \sigma}{3 \kappa \rho} ,
\end{equation}
\begin{equation}
\xi_{\text{rad}} = \frac{\gamma - 1}{\gamma} \frac{T}{P} \chi,
\end{equation}
where $\sigma$ is the Stefan-Boltzmann constant and $\kappa$ is the radiative opacity. We have interpolated the Rosseland opacity $\kappa(\rho, T)$ from the OPAL table for solar composition \citep{IglesiasRogers1996}. The kinematic viscosity is taken from \citet{Spitzer1962}, $\nu_d \simeq 21$ cm$^2$ s$^{-1}$. The radiative viscosity is given by e.g. GS67 and yields a small contribution to the total viscosity; with our solar model and the Rosseland opacity interpolated from the OPAL table for solar composition, we determine $\nu_r \simeq 2$ cm$^2$ s$^{-1}$. We adopt the same value of MBS04 for the resistivity $\eta$. Finally, the values we adopt for the diffusivities at $r = 0.70 R_\odot$ are: $\xi_{rad\odot} = 1.4 \times 10^7$ cm$^2$ s$^{-1}$, $\nu_\odot \simeq 23$ cm$^2$ s$^{-1}$, $\eta_\odot = 596$ cm$^2$ s$^{-1}$.

Since $\partial_z \Omega \ne 0$, the naive expectation is that the GSF instability should be present.   In what follows, we discuss the occurrence of this instability in the upper radiative zone of the Sun in section \ref{sec:GS}. We then present in section \ref{sec:trials} the evolution of a variety of non-axisymmetric solutions of equations \eqref{Eq1} - \eqref{Eq5}.

\subsection{The magnetic field in the radiative zone of the Sun}

If a magnetic field with finite poloidal components $B_R$, $B_z$ and finite derivatives of angular velocity $\partial_R \Omega$, $\partial_z \Omega$ are also present, the azimuthal component of the background 
field $\bb B$ increases linearly with time.   This of course is also encountered in classical accretion disc theory of
the magnetorotational instability (MRI).  It makes no difference to the analysis so long as the background field remains weak.   In disc theory, the rapid breakdown of the flow into MHD turbulence renders moot the problem of a progressive linear build-up of the azimuthal field.   
In the stellar case, this could be an issue in principle for an ostensibly stable rotation profile.   Of course this is true for {\em any} model analysis that does not have $\bb{B\cdot\nabla}\Omega=0$, and stability issues can arise  (see e.g. \citealt{Braithwaite2009}).  The Sun is, however, not in a state of uniform rotation, and it is magnetised. After 4.5 billion years it is not unreasonable to assume that the magnetic field lies very nearly in constant $\Omega$ surfaces. Since our findings of local stability are not sensitive to field geometry, we will ignore the problem of strong field build-up, assuming that in regions of the Sun that are not actively turbulent and dissipating magnetic field, such strong growth is not occurring.

\section{Axisymmetric perturbations: the Goldreich-Schubert-Fricke instability in the upper radiative zone of the Sun} \label{sec:GS}

The GSF instability affects axisymmetric perturbations in a rotating medium if the angular velocity is not constant on cylinders, i.e. $\partial_z \Omega \ne 0$, and a finite thermal diffusion coefficient $\xi_{\text{rad}}$ is present. We study here the evolution of perturbations under such conditions, allowing for the introduction of a finite viscosity $\nu$.

Analytically, the unstable modes may be identified by determining the sign of the last term of the dispersion relation of GS67, eq. (32): whenever it is negative, the equation has an unstable solution. In the $\nu = 0$ case, it is easy to see that unstable modes will be found whenever $\partial_z \Omega \ne 0$, independently of the value of $\xi_{\text{rad}}$. 

\begin{figure}
	\centering
	\includegraphics[width=0.5\textwidth, clip=true, trim=0cm 0cm 0cm 0cm]{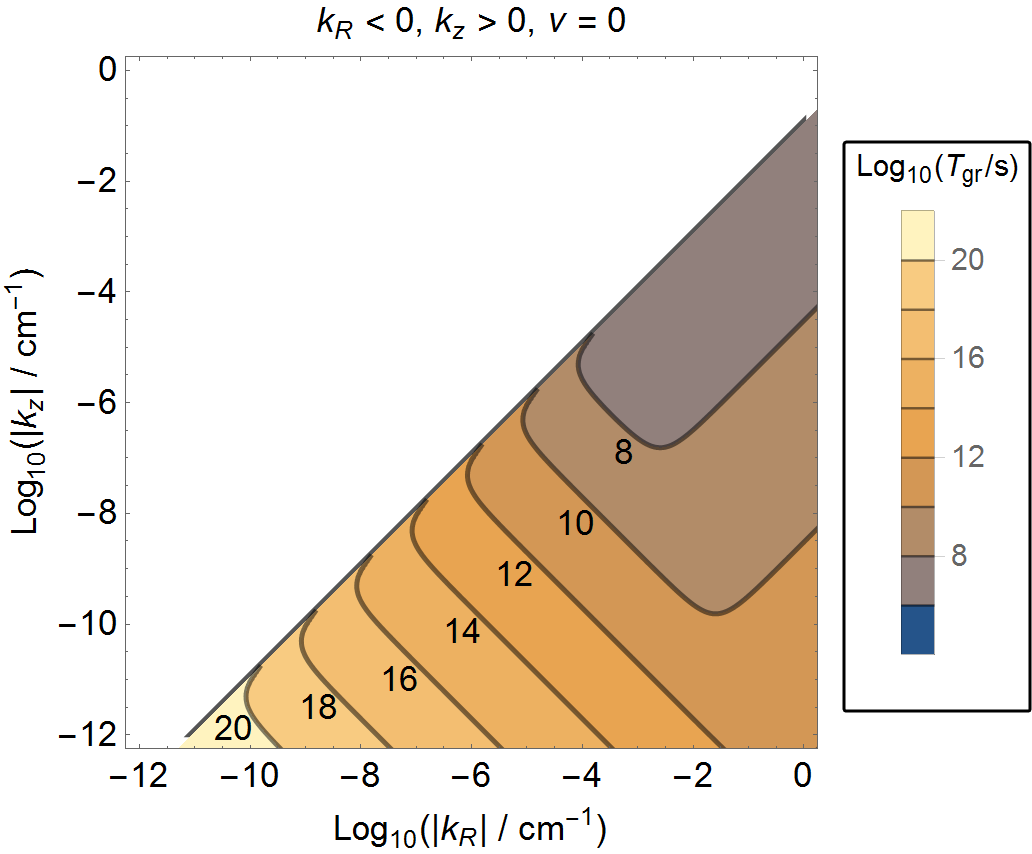}					
	\caption{\label{figGS0bis}Region of the $k_R < 0, k_z > 0$ plane where the GSF axisymmetric instability occurs in the Sun, for $r = 0.7 R_\odot$, $\theta = 45 \degree$, in case of no viscosity, for an angular velocity pattern corresponding to model A. The growth time-scale $T_{\text{gr}}$ of the instability is also shown. The numbers next to the iso-contours correspond to the value of $Log_{10}(T_{\text{gr}})$ with $T_{\text{gr}}$ expressed in seconds.}
\end{figure}

The region of the $k_R - k_z$ space corresponding to the unstable models is, by the classical GS criterion,
\begin{equation}
{\partial l\over \partial R} - {k_R\over k_z}{\partial l\over \partial z} > 0 ,
\end{equation}
where $l=R^2\Omega$, the specific angular momentum.
For $r = 0.7 R_\odot$, $\theta = 45 \degree$, there are unstable modes only when $k_R$ and $k_z$ have opposite signs.
We show in figure \ref{figGS0bis} the region of the $k_R < 0, k_z > 0$ plane where unstable modes exist, for a large range of values of $k_R$ and $k_z$, for an angular velocity pattern corresponding to model A. We also report the growth rate $T_{\text{gr}}$ of the axisymmetric perturbations, derived from the dispersion relation by GS67. 

Matters are quite different when a finite viscosity is taken into account. As discussed in section \ref{sec:intr1}, some authors noted that even a small viscosity $\nu \ll \xi_{\text{rad}}$ has a stabilising effect on the GSF instability. We have solved the GS67 axisymmetric dispersion relation (their equation (32)).  As the equation is quite lengthy, we shall not reproduce it here, but refer the reader directly to the original paper.  Over the range of $k_R, k_z$ shown in figure \ref{figGS0bis} with $\nu = \nu_\odot$, we have found no unstable behaviour: \emph{all the axisymmetric modes appear to be stable}. 

Performing the same analysis for values of the co-latitude $\theta$ in the $5 \degree - 85 \degree$ range, we find stability for all $\theta$.  We have also performed the same analysis for the angular velocity pattern corresponding to model B, which is numerically similar to model A, and find very little behavioural difference for $\nu = 0$ and essentially no difference for $\nu = \nu_\odot$.  Finally, we have explored the case $\nu = 0.5 \nu_\odot$, to determine whether an order unity uncertainty on the value of $\nu$ would bear any consequence on our analysis. Once again, even with this reduced value of the viscosity there are no unstable modes.  Evidently, the axisymmetric GSF instability does not occur in the Sun, even if its pattern of rotation features a moderate gradient of angular velocity, compatible with the data from helioseismology.   

Although the Sun's rotation profile does not appear to be vulnerable to the axisymmetric GSF instability, conditions in other stars may be, and it is of interest to pursue equations \eqref{Eq1} - \eqref{Eq5} to study the behaviour of a fluid which is axisymmetrically unstable in the presence of finite viscosity, allowing for the presence of a magnetic field and finite $k_\phi$. The analysis of the current paper is limited to the stability of modes in the current Sun, but a more general study of the onset of the GSF instability in other environments will be investigated in a subsequent publication.

\section{Stability of the upper radiative zone of the Sun} \label{sec:trials}
The main point of this paper is the study of the local stability of {\em non-axisymmetric displacements} in a differentially rotating background, which requires the solution of a set of ordinary differential equations not reducible to
an algebraic dispersion relation.   Since the wavenumber is varying, this approach seems more prudent than trying to incorporate an intrinsically time-dependent wavenumber into a dispersion relation formalism (e.g., \citealt{KaganWheeler2014}). 

As seen from equations \eqref{kRoft} - \eqref{kzoft}, perturbations with $k_\phi \ne 0$ become asymptotically axisymmetric as $t\rightarrow \infty$, provided that $\del \Omega \ne 0$, though the time dependence of
$\bb{k}$ is not lost. 
A general stability analysis of the models of rotation of the upper radiative zone described in section \ref{sec:stabilityofthesun} requires the full solution of equations \eqref{Eq1} - \eqref{Eq5} for any given initial wave vector $\bb k_0$, initial conditions $\delta \bb v_0$, $\delta \rho_0$, and $\delta \bb B_0$. Non-modal 
(in the sense of non-exponential and non-trigonometric) problems with similar features have recently been discussed for the onset of the MRI in discs.   Here, transient growths of (eventually) stable modes are of particular significance. \citet{Squire2014} discuss this topic and the relevant methods in some detail.   (See \citealt{Trefethen} for a rigorous mathematical presentation.)   This method is computationally demanding and lies outside the scope of the current paper.   

Here, we present a study of the transient evolution of a variety modes for an arbitrary set of initial conditions, both systematically and randomly selected.   Our aim is to identify regions of instability (more accurately, large transient growth) in the $k_R - k_\phi - k_z$ space, if any exist.   As we have seen, the GSF instability occurs in a broad section of the $k_R - k_z$ plane when viscosity is ignored.   It is in principle possible that nonaxisymmetry might also tip the balance.     In fact, we find no large transient growth.   We caution that {\em the limitations of our approach do not allow us to make a definitive claim for the absolute stability of the models,} as we have not performed a complete exhaustive study of parameters.   However, if these models are unstable, something more than simple GSF behaviour is involved.

\subsection{Details of our modes selection}
We have selected a single set of initial conditions as follows: we set $\delta v_z = 1$ cm s$^{-1}$ (the numerical value is arbitrary),  $\delta v_\phi = 0$, $\delta \rho_0 = 0$, $\delta \bb B_0 = 0$.  We assign to $\delta v_R$ a different value for each mode, chosen as to satisfy equation \eqref{Boussinesq2}. We have studied modes with initial wave vector components in the range:
\begin{equation}  \label{krange}
k_{R0}, k_\phi, k_{z0}: \ \pm \frac{2 \pi}{10^{-2} R_\odot} \rightarrow \pm \frac{2 \pi}{10^{-14} R_\odot},
\end{equation}
The first limit guarantees that the wavelength of the perturbation is small compared to the typical scale height of the structural properties of the Sun ($\sim 10^{-1} R_\odot$), as required by the WKB approximation. The second limit guarantees that the wavelength is large compared to the mean free path of the particles in the radiative zone of the Sun (MBS04). 

We focus on the non-degenerate case in which the values of $k_{R0}, k_\phi$, and $k_{z0}$ are such that the absolute value of the ratio of any two of the wave vector components is not smaller than $10^{-3}$.  As a consequence, we cannot reproduce instabilities that would occur in the extreme degenerate case; but the  axisymmetric limit can be addressed via the dispersion relation.   We select a value for the total integration time by the following considerations.  From \eqref{kRoft} - \eqref{kzoft}, $k_R$ and $k_z$ depend on time, while $k_\phi$ is constant, so that the behaviour of any perturbation becomes increasingly (large wavenumber) axisymmetric with time. The conditions for the perturbation to be approximately axisymmetric are:
\begin{equation}  
R k_{\phi} \frac{\partial \Omega}{\partial R} t \gg k_{R0}, \qquad R k_{\phi} \frac{\partial \Omega}{\partial R} t \gg k_{\phi},
\end{equation}
\begin{equation}  
R k_{\phi} \frac{\partial \Omega}{\partial z} t \gg k_{z0}, \qquad R k_{\phi} \frac{\partial \Omega}{\partial z} t \gg k_{\phi}.
\end{equation}
With values of the derivatives of $\Omega$ typical of models A and B, and the requirement that $|k_\phi / k_{R0}|$ and  $|k_\phi / k_{z0}|$ are not smaller than $10^{-3}$, these conditions are satisfied if $t >> 10^9$ s $\cong$ 30 y. We follow the evolution of the perturbations an order of magnitude beyond this, after which time the perturbations are very close to axisymmetric.

To solve equations \eqref{Eq1} - \eqref{Eq5}, it is necessary to specify the strength and geometry of the background magnetic field in the upper radiative zone. We note that $\bb B$ appears in equations \eqref{Eq1} - \eqref{Eq5} only via the term $\bb k \bcdot \bb v_A$. We may thus treat this term as a constant parameter. Unfortunately, $\bb k \bcdot \bb v_A$ is effectively a natural frequency of the system and, in the most interesting case in which it is comparable to (or higher than) the angular velocity $\Omega$, its presence makes following the evolution of the system computationally more demanding.  Both for this reason and because the magnetic field appears not to play a destabilising role, we focus on the hydrodynamic case, introducing a magnetic field in a small number of runs described in section \ref{sec:magnetic}.

\subsubsection{Systematic approach}

We study the evolution of a sample of $\sim 10^4$ modes for both model A and B. The wave vector components are given by the angles $\theta_k$ and $\phi_k$ describing the orientation of $\bb k$ in spherical wavenumber space: 
\begin{equation}  
k_{R0} = |\bb k| \sin(\theta_k) \cos(\phi_k), 
\end{equation}
\begin{equation}  
k_\phi = |\bb k| \sin(\theta_k) \sin(\phi_k), 
\end{equation}
\begin{equation}  
k_{z0} = |\bb k| \cos(\theta_k) .
\end{equation}
We follow perturbations at $r = 0.70$ R$_\odot$ for a variety of modes selected as follows. We choose 5 equally spaced values of the co-latitude $\theta$ in the interval $[5 \degree, 85 \degree]$; we assign values of $|\bb k|$ so that $\log_{10}(|\bb k|) = -10, -9, ..., 3$; we select 5 equally spaced values of $\theta_k$ in $(0 \degree, 180 \degree)$, and 10 equally spaced values of $\phi_k$ in $(0 \degree, 360 \degree)$. We also include all the relevant values of $\theta_k$ and $\phi_k$ that are $1 \degree$ away from $0 \degree, 90 \degree, 180 \degree, 270 \degree$.   For present purposes, a transient growth coefficient has been defined as the maximum of $\delta v^2(t) / \delta v^2(t = 0)$. For both models, this analysis has not revealed any mode that is unstable or shows significant transient growth. 

\subsubsection{Random selection - hydrodynamic case} \label{sec:nonmagnetic}
In the $\bb v_A = 0$ case, we have selected $10^5$ runs for each of 9 equally spaced values of the co-latitude $\theta$ in the interval $5 \degree - 85 \degree$, for both models A and B, for a total of $1.8 \times 10^6$ modes, and tracked the evolution of these modes. The modes have been generated from uniform distributions for $\log_{10}(|k_R|)$, $\log_{10}(|k_\phi|)$, $\log_{10}(|k_z|)$ with the additional constraint that the absolute value of the ratio of any two of the wave vector components is not smaller than $10^{-3}$.

\begin{figure}
	\centering
	\includegraphics[width=0.5\textwidth, clip=true, trim=0cm 0cm 0cm 0cm]{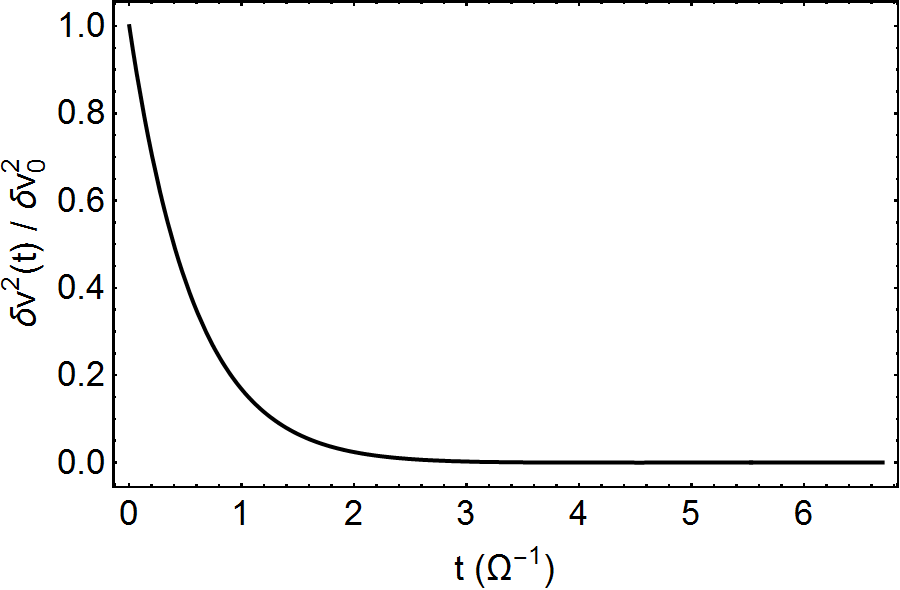}					
	\caption{\label{figTypical}Typical behaviour of a perturbation. This particular solution was generated for the model A at $\theta = 45 \degree$ with wave vector components $k_{R0} = 2.1 \times 10^{-4}$ cm$^{-1}$, $k_\phi = 4.2 \times 10^{-7}$ cm$^{-1}$, and $k_{z0} = 1.3 \times 10^{-5}$ cm$^{-1}$. The temporal axis is expressed in units of $\Omega^{-1} = 3.77 \cdot 10^5$ s, about 4.4 days.}
\end{figure}

\begin{figure}
	\centering
	\includegraphics[width=0.5\textwidth, clip=true, trim=0cm 0cm 0cm 0cm]{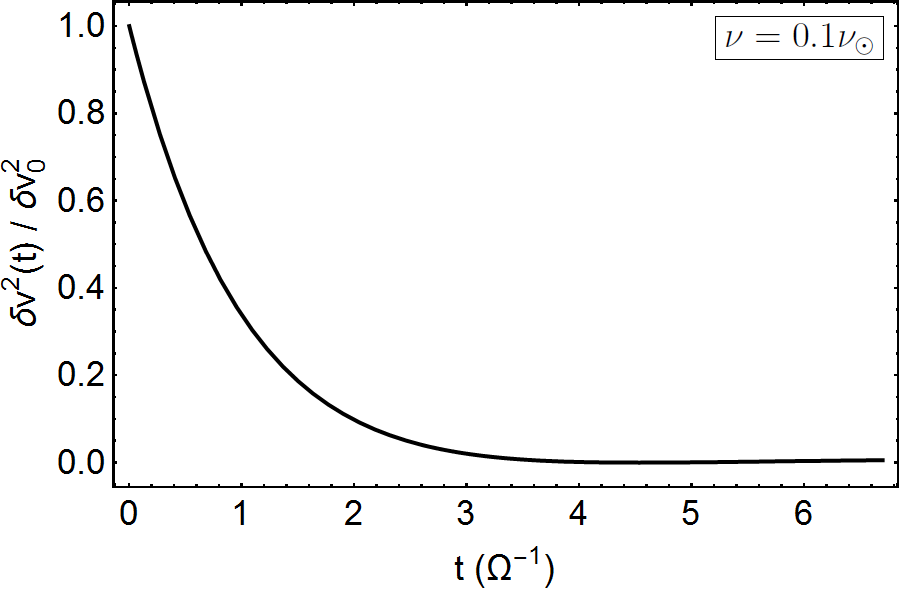}					
	\caption{\label{figReducedViscosity}Behaviour of the same perturbation of figure \ref{figTypical} with $\nu = 0.1 \nu_\odot$. The units are as in figure \ref{figTypical}.}
\end{figure}

No unstable modes were found. The vast majority of modes were damped, with no traceable growth.
A typical evolution is shown in figure \ref{figTypical}. The time-scale of the damping and, when present, the oscillations, vary significantly between different modes. The damping of the perturbations appears to be predominantly viscous.    Figure \ref{figReducedViscosity}, for example, shows the behaviour of the perturbation with the same wave vector as that of figure \ref{figTypical}, but with an artificially reduced viscosity of $\nu = 0.1 \nu_\odot$.  

\begin{figure}
	\centering
	\includegraphics[width=0.5\textwidth, clip=true, trim=0cm 0cm 0cm 0cm]{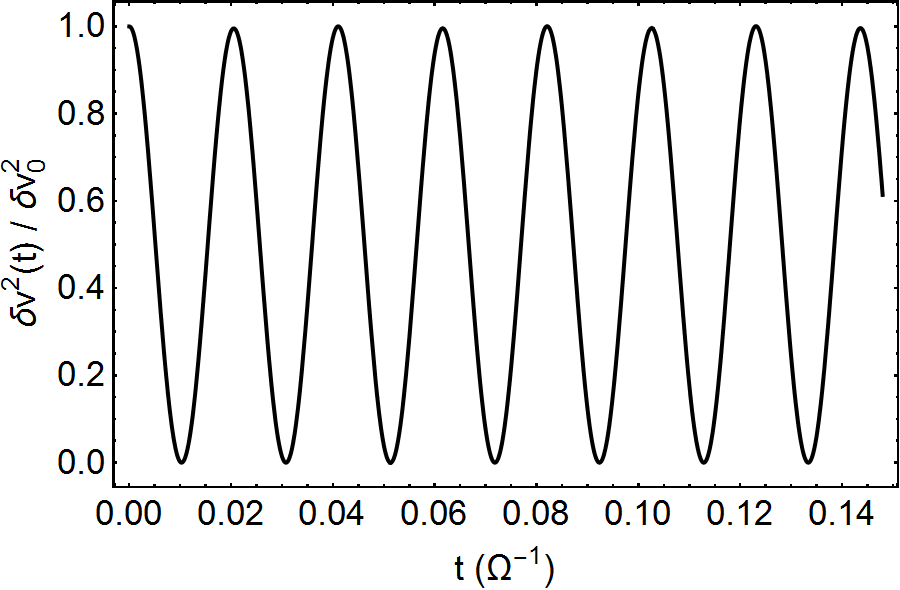}					
	\caption{\label{figTransient}Very short time-scale oscillations of a perturbation featuring 
{\em long term} transient growth (figure \ref{figTransient2} below). This particular solution was generated for the model A at $\theta = 85 \degree$ and has wave vector components $k_{R0} = 8.7 \times 10^{-9}$ cm$^{-1}$, $k_\phi = 1.5 \times 10^{-10}$ cm$^{-1}$, and $k_{z0} = 4.9 \times 10^{-9}$ cm$^{-1}$.  The units are as in figure \ref{figTypical}. The black line shows $\delta v^2(t)/\delta v_0^2$.}
\end{figure}

\begin{figure}
	\centering
	\includegraphics[width=0.5\textwidth, clip=true, trim=0cm 0cm 0cm 0cm]{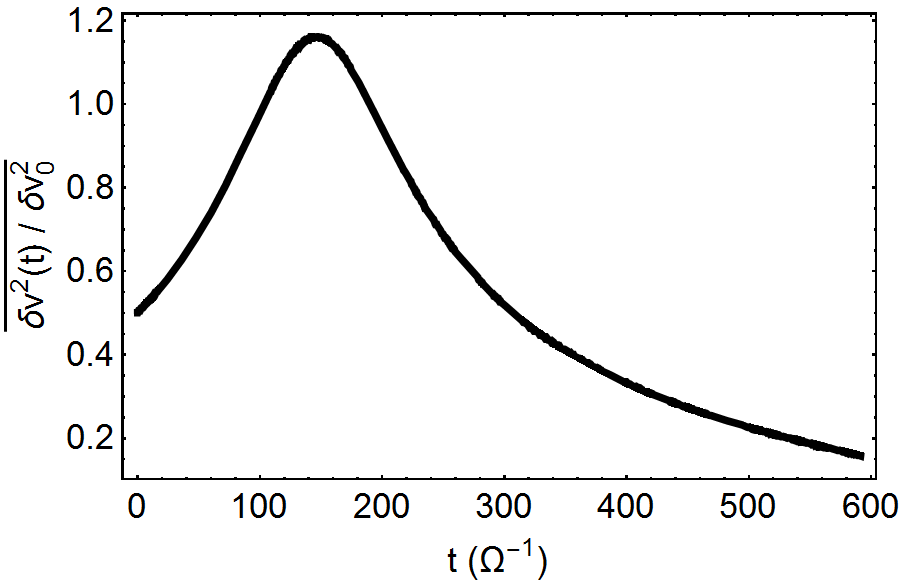}					
	\caption{\label{figTransient2}Integral average of $\delta v^2(t)/\delta v_0^2$ over a time of $\Omega^{-1}$ for the perturbation of figure \ref{figTransient}. The maximum growth factor is approximately 2.4 and is reached at the peak of the short time-scale oscillations near $t \sim 140 \Omega^{-1}$.}
\end{figure}

We have identified about $10^2$ modes that present a moderate transient growth before eventually damping to zero. In all these cases, the maximum growth factor is of order unity. These modes have typically small wave number, $|\bb k| \lesssim 10^{-8}$ cm$^{-1}$, feature multiple time-scales behaviour, and have a much longer damping time-scale than the others. We show one such displacement in figures \ref{figTransient} (the short time-scale oscillations) and \ref{figTransient2} (the long term integral average of $\delta v^2(t) / \delta v_0^2$). 


\subsubsection{Magnetic case} \label{sec:magnetic}
As noted above, $\bb B$ appears in equations \eqref{Eq1} - \eqref{Eq5} only via the term $\bb k \bcdot \bb v_A$, a natural frequency of the system and a constant parameter. Since our focus is on the stability of the differential rotation patterns, we examine a variety of cases with $\bb k \bcdot \bb v_A \ll \Omega$, $\bb k \bcdot \bb v_A \sim \Omega$, or $\bb k \bcdot \bb v_A \gg \Omega$, studying the interplay between the magnetic and rotational effects. 

We selected $10^4$ random modes for each of 9 equally spaced values of the co-latitude $\theta$ in the interval $5 \degree - 85 \degree$, for both models A and B, for a total of $1.8 \times 10^5$ modes. As in section \ref{sec:nonmagnetic}, we varied $k_{R0}$, $k_\phi$, and $k_{z0}$ in the range \eqref{krange}. Adopting a uniform distribution for $\log_{10}(|\bb k \bcdot \bb v_A|)$, we independently assigned random values to $\bb k \bcdot \bb v_A$ in the range:
\begin{equation}
\bb k \bcdot \bb v_A: \ \pm 10^{-2} \Omega \rightarrow \pm 10^{2} \Omega .
\end{equation}

As before, we found no unstable behaviour.   The vast majority of modes being damped with no traceable growth. We have identified about $10$ modes that present a transient growth before being eventually damped to zero, with maximum growth factor of order unity.

\begin{figure}
	\centering
	\includegraphics[width=0.5\textwidth, clip=true, trim=0cm 0cm 0cm 0cm]{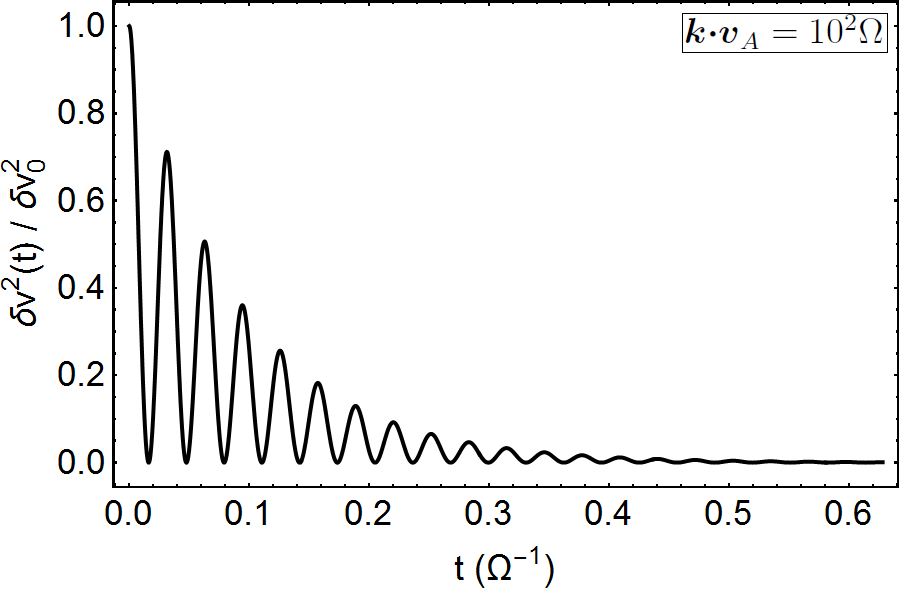}					
	\caption{\label{figMagneticNormal}Behaviour of the same perturbation of figure \ref{figTypical} in the $\bb k \bcdot \bb v_A = 10^{2} \Omega$ case. The units are as in figure \ref{figTypical}.}
\end{figure}

\begin{figure}
	\centering
	\includegraphics[width=0.5\textwidth, clip=true, trim=0cm 0cm 0cm 0cm]{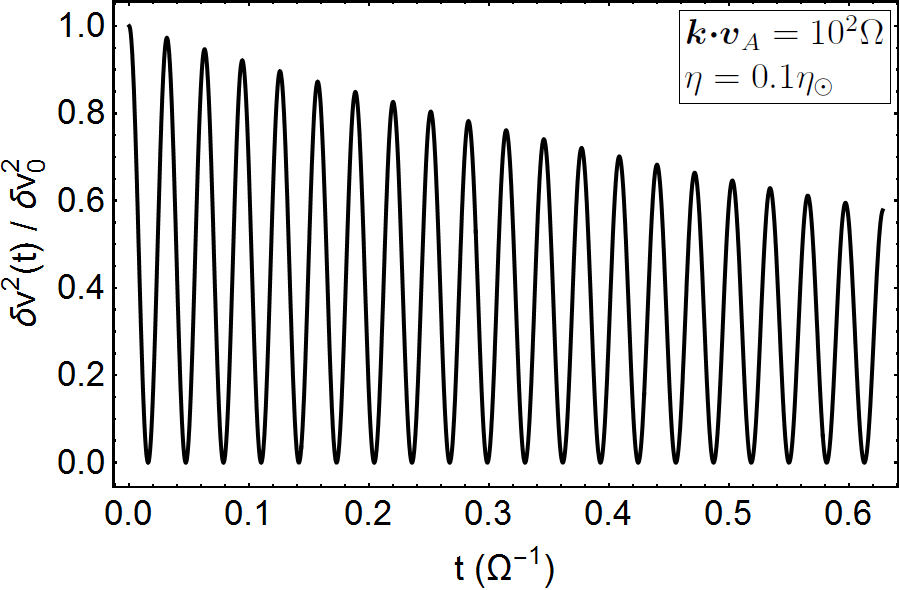}					
	\caption{\label{figMagneticReducedResistivity}Behaviour of the same perturbation of figure \ref{figTypical} in the $\bb k \bcdot \bb v_A = 10^{2} \Omega$, $\eta = 0.1 \eta_\odot$ case. The units are as in figure \ref{figTypical}.}
\end{figure}

High values of $\bb k \bcdot \bb v_A$ cause short time-scale oscillations of the perturbations, which are ultimately damped. Figure \ref{figMagneticNormal} shows the behaviour of a perturbation with the same wave vector used in figure \ref{figTypical},but  with $\bb k \bcdot \bb v_A = 10^{2} \Omega$. As might have been expected, given that $\eta_\odot \gg \nu_\odot$, in the case $\bb k \bcdot \bb v_A \gg \Omega$ the damping is primarily due to the resistivity, not the viscosity, and it occurs on a shorter time-scale. While decreasing $\nu$ has a minor effect on the results of figure \ref{figMagneticNormal}, decreasing $\eta$ lessens the amplitude decrease.   For the selected rotation profiles, the effect of the magnetic field, in contrast to MRI vulnerable systems, is to render the system yet more stable. Figure \ref{figMagneticReducedResistivity} shows the behaviour of a perturbation with the same wave vector, with magnetic parameter $\bb k \bcdot \bb v_A = 10^{2} \Omega$, and $\eta = 0.1 \eta_\odot$.

\section{Conclusions and Limitations}
The local stability of a differentially rotating medium is a subject of great importance in the study of astrophysical fluids, most notably in the fields of accretion discs and rotating stars. The effects of non-axisymmetry, diffusive processes, and magnetic fields are often subtle, but can be critical in many applications. We have presented here a very general analysis, using the local linearised equations of a weakly magnetized, differentially rotating fluid, with finite thermal conductivity, viscosity, and resistivity for non-axisymmetric WKB displacements.    Locally
co-moving Lagrangian coordinates have been employed.   In the non-axisymmetric case, the equations are non-modal and the problem cannot be reduced to a dispersion relation.

Differential rotation stratified on cylinders is often a favoured model profile for the interiors of stars, discs, and, increasingly often, planets.  This is of course inevitable when a barotropic equation of state is used, but it is often justified on the basis of the classical axisymmetric Goldreich-Schubert criterion.  Under conditions valid in the radiative zone of the Sun, our analysis shows that this is misleading, and the Sun evidently has little difficulty 
maintaining a GSF-violating profile.    Realistic values of $\nu$ and baroclinic rotation models compatible with the data from helioseismology prevent the growth of the GSF instability.   It should  be noted that, while nominally tailored for solar rotation, the GSF study was carried out long before the rotation profile of the Sun was actually known.   The rotation-on-cylinders stability criterion applies to much stronger differential shear (which was the target of investigation of those authors) than is actually present in the solar interior.  

We have calculated the evolution of non-axisymmetric displacements in the upper radiative zone and the tachocline of the Sun, near $r = 0.7 R_\odot$ at various latitudes.   We have found neither unstable disturbances nor strong transient growth.  Patterns of solar rotation featuring angular velocity gradients similar to those inferred from the helioseismology data are more stable than is commonly realised, though to be sure more work needs to be done to address questions of global stability.  The investigation described here was carried out on the rotation pattern proposed by CBP15, which is in strict, static radiative equilibrium. The feasibility of this model may alleviate the problem of the circulation-induced spreading of the tachocline \citep{SpiegelZahn1992}. 

We re-emphasise that only local stability has been discussed here. 
Moreover, nonlinear destabilising shear processes may be present, either locally or on global scales  \citep{Zahn1975}. While a full understanding of this remains elusive \citep{MenouLeMer2006}, it is currently considered of great importance for the transfer of angular momentum in stars.

Finally, our analysis assumes a uniform chemical composition in the star.   Although this is accurate in the bulk of the radiative zone, other venues and other problems might benefit from relaxing this hypothesis. If the tachocline of the Sun is characterised by strong compositional gradients, for example, this would aid in confining the magnetic field of the radiative zone (see \citealt{Christensen2007}, \citealt{Wood2011}).

\section*{Acknowledgements}
AC acknowledges support from the University of Oxford. SAB acknowledges support from the Royal Society in the form of a Wolfson Research Merit Award. We would like to thank the anonymous referee for carefully reading our manuscript and giving in-depth comments which substantially helped improving the quality of the paper.

\bibliography{References}
\bibdata{References}

\label{lastpage}

\end{document}